# Ultra-Thin Metal Films for Enhanced Solar Absorption

**N.Ahmad, J.Stokes, N.A.Fox[1], M.Teng and M.J.Cryan**
*Department of Electrical and Electronic Engineering and [1]School of Physics, University of Bristol*
*Nathan.Ahmad@bristol.ac.uk, M.Cryan@bristol.ac.uk*

**Abstract**: This paper presents modelled results for optical absorption in ultra-thin films of nickel, gold and silver across the solar spectrum. It is found in the case of nickel there is an optimum thickness for maximum solar absorption around 10-13nm. It is believed that this is a result of the real and imaginary parts of its refractive index being of similar magnitude across the solar spectrum which can give rise to very strong thin film absorption.

## 1. Introduction

There are a number of alternative approaches to solar photovoltaics that are being pursued as low cost, efficient means of generating electricity. These include solar thermal technologies [1], where solar energy is used to directly heat liquids in order to drive turbines and solar thermionics [2], where solar energy heats a metal in vacuum such that electrons are emitted into the vacuum and collected by a cooled anode to create an electric current [3]. Heat based solar technologies have the major advantages that the absorbed wavelengths are not restricted by a semiconductor band gap and are often much simpler in terms of device fabrication since they do not require expensive semiconductor growth. As with all solar technologies it is critical to absorb as much of the incident solar energy as possible which implies that the reflectance should be minimised across the whole of the solar spectrum. Heat based technologies have the added requirement to maximise temperature rise, which implies minimising radiation in the thermal infrared band. Thus there are conflicting requirements of strong absorption in the visible, with minimum emission in the far infrared range; the figure of merit for this is termed selectivity. There are a number of approaches that have been pursued to meet these requirements, notably cermets [4] which are composite materials made from ceramic and metal nanoparticles formed on a strongly reflecting substrate. These are designed such the metal particles absorb strongly in the visible, but transmit strongly in the far infrared, the reflecting substrate then ensures maximum reflection in the infrared which results in minimum emission of heat radiation. There are a number of recent new approaches to this problem which involve surface micro and nano-structuring, including photonic crystals [5], nanoantennas [6], metamaterials [7] and plasmonics [8,17]. These techniques can strongly modify the emission and absorption properties of materials and have the potential to produce low cost, high performance absorbing surfaces when combined with low cost micro and nano fabrication techniques such as 2D and 3D printing and NanoImprint Lithography(NIL). An important first step prior to the design of the surface structure is to determine which material or materials are optimum for a particular application. For solar thermal applications good high temperature performance is critical and in solar thermionic applications temperatures of 1100-1300K are expected which is beyond the capability of many conventional cermets and black absorbing paints. The advantage of 2D or planar surface structuring is that only one material is required and the thermal properties of the surface can be very close to that of the bulk material. In more complex 3D surfaces this can be more difficult, but free standing single material nano and microscale structures can be fabricated using 3D printing and laser writing techniques followed by back filling or electroplating [9]. This paper explores the fundamental absorbing properties of a number of metals which could serve as a base material for a high temperature, high selectivity structured surfaces and explores the properties of ultra-thin films in terms of reflection, transmission and absorption across the solar spectrum. These films have the potential to be formed into multilayer structures such as woodpile [5] and a detailed understanding of the performance of a single layer is required before moving to more complex 3D structures.

The optical properties of metals have been exploited for thousands of years with the prominent example being silvered glass mirrors. There are less obvious examples such as coloured glasses which take their colour from plasmon resonances in metallic nanoparticles [10]. The main optical property of metals is their high reflectivity, with some of the highest values in the visible being observed silver. Many of these properties can be strongly dependent on surface morphology with roughened surfaces producing more lustrous or matt finishes. There are some metals such as nickel which have naturally lower reflectivity in the visible and this makes them good candidates for solar absorbers. Allied to the optical properties, for solar thermal applications, good high temperature performance is required and this can be summarised by the requirement for a high melting point. The highest melting point metal is tungsten (3422°C) and hence it is used in lighting applications in filament bulbs. This high melting point can make it more difficult to be deposited on to surfaces for planar applications such as solar thermionics and nickel is found to be a good compromise with a melting point of 1453°C. Thus this paper will focus on the properties of thin films of nickel and compare them to more the widely used metals: gold and silver.

There has been much work carried out on absorption in thin lossy media with some recent examples and references therein giving a good overview [16,17]. It has been shown [17] for example that for a media with real part of the refractive index, N, and imaginary part k, that when N≈k optimum absorption can occur in ultrathin films. It has also been shown using an optical impedance model [16] that there is an optimum thickness for obtaining maximum absorption. In this paper we will look in detail at the performance of nickel, gold and silver and show that nickel can give excellent thin film absorption over the entire solar spectrum.

Much research has also been carried out on ultra-thin metal films for use as transparent conducting layers in numerous optical and thermal applications [11,12]. These investigations have focused mostly on the transmission of thin films and less on absorption, but we are able to compare some of our results with experimental data in order to validate the use of our refractive index data in the case of ultra thin films.

The paper will begin with a study of the bulk properties of these metals and then use the well known Transfer Matrix Method to calculate the optical properties of thin layers. In the case of nickel an optimum thickness of 12.5nm is found for maximum absorption across the solar spectrum. This optimum is not observed for gold or silver where thicker layers will always produce higher absorption, which asymptotically approaches a maximum at infinite thickness. This is interesting for a number of reasons in particular that less material can be used whilst maintaining high absorption.

## 2. Results

One of the fundamental electromagnetic properties of a material is its relative dielectric constant, $\varepsilon_r$ which in general can be complex, $\varepsilon_r = \varepsilon_{rre} + j \varepsilon_{rim}$. Where the real part is related to the phase shift that occurs on propagation through the material and the imaginary part represents absorption effects such as intraband absorption [13]. Figure 1 below shows the relative dielectric constant data for nickel, gold and silver across the solar spectrum [14] (note that the negative is the real part is being plotted, giving rise to positive numbers on the graphs). We are using 300nm-2500nm which covers the majority of the solar spectrum. Figure 1 shows one of the unusual properties of some metals in the visible range: they possess a negative real part to the dielectric constant. This gives rise to effects such as surface plasmons [8,17] under certain conditions which can dramatically enhance absorption. In this paper we study only normal incidence on flat surface and in this case surface plasmons cannot be generated, but in structured surfaces they will play a key role and so this should be considered carefully when selecting a material. The figure also shows that nickel has much higher absorption across the solar spectrum while silver has very low absorption, resulting in its highly reflective surface. Gold has strong absorption towards the blue region of the visible spectrum giving it its classic colour.

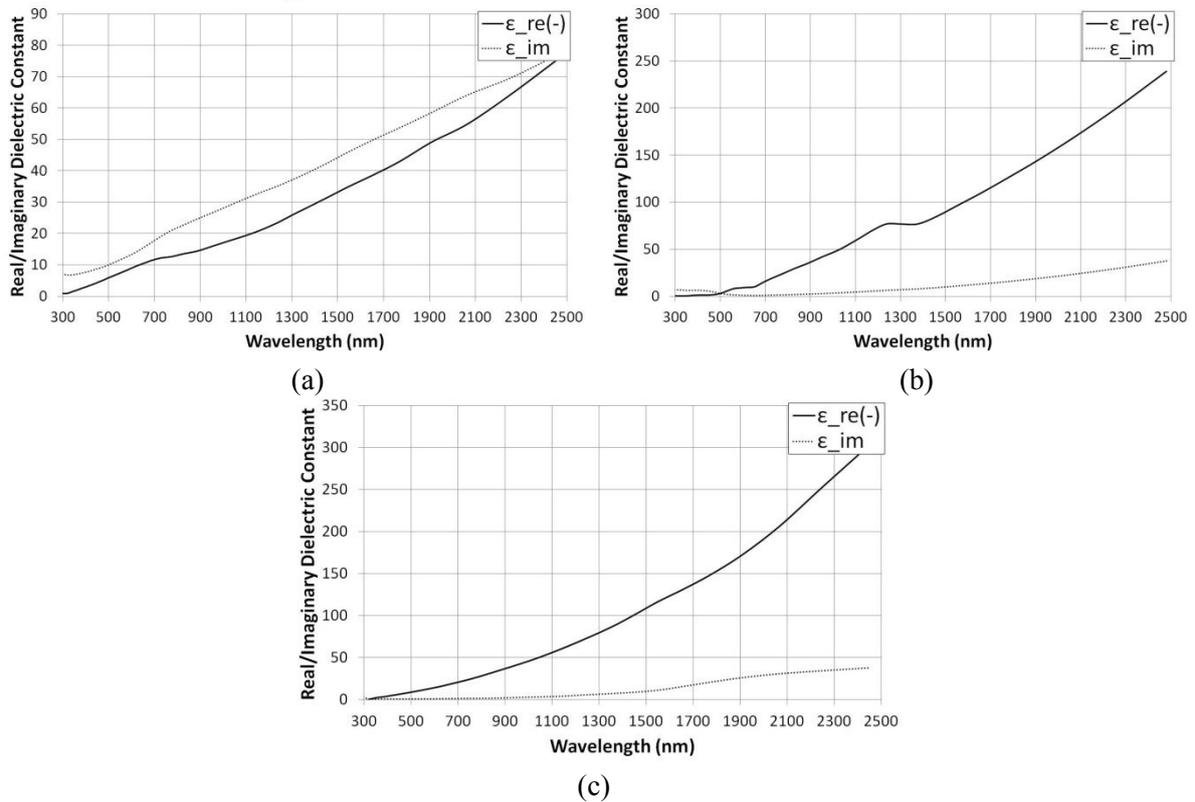

Figure 1: Real (negative) and Imaginary part of the relative dielectric constant of (a) nickel, (b) gold and (c) silver [14]

In optics it is often more straightforward to use complex refractive index, $n=N+jk$ as opposed to dielectric constant and $N$ can be derived from $\varepsilon_r$ in a straightforward way as shown in [18] for example. The refractive index data is shown in Figure 2

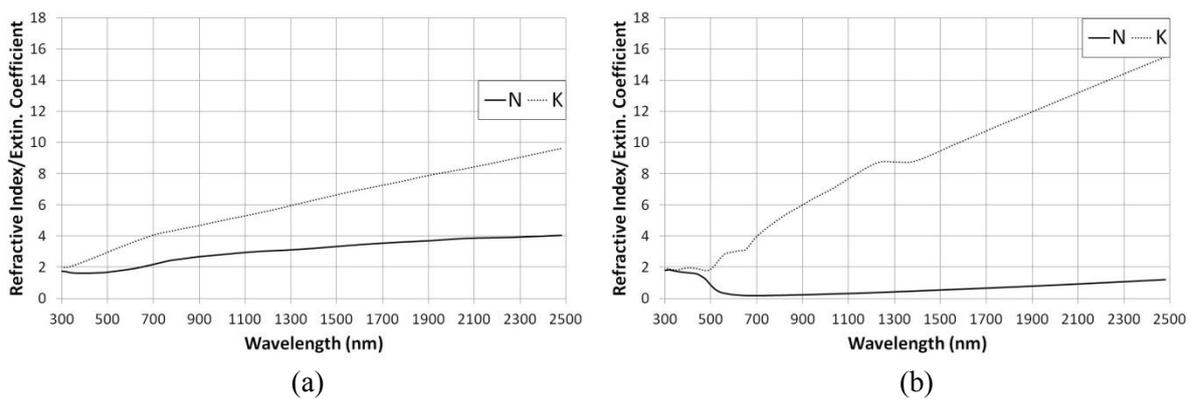

(a)          (b)

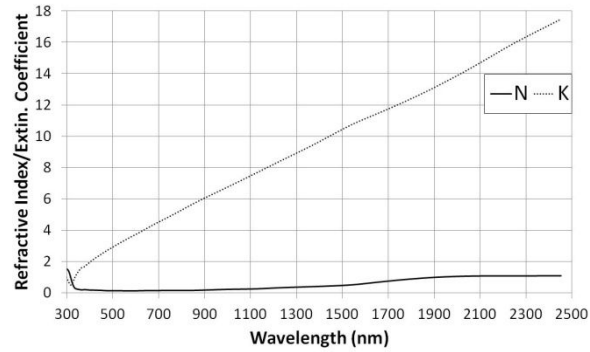

(c)

Figure 2: Real and imaginary parts of the refractive indices for the metals (a) nickel, (b) gold and (c) silver

Figure 2 shows that for the noble metals, gold and silver, the imaginary part of n is much bigger than the real part, which in fact can be less than unity due to the negative real part of the dielectric constant. Whereas for nickel the real and imaginary parts are of similar magnitudes and is thus close to fulfilling the optimum conditions for thin film absorption mentioned above [17]. This paper will now analyse reflection, transmission and absorption and determine the optimum thickness for wideband solar absorption for nickel and compare this with the cases of gold and silver.

This refractive index data can now be used to calculate the bulk reflectance, R at an air-metal interface given which can be calculated using the well known formula:

$$R = \left|\frac{N+jk-1}{N+jk+1}\right|^2 \qquad (1)$$

The results are shown in Figure 3. These results show as expected that gold and silver are strongly reflecting across most of the solar spectrum and the more matt surface of nickel results in stronger absorption and hence lower reflectance.

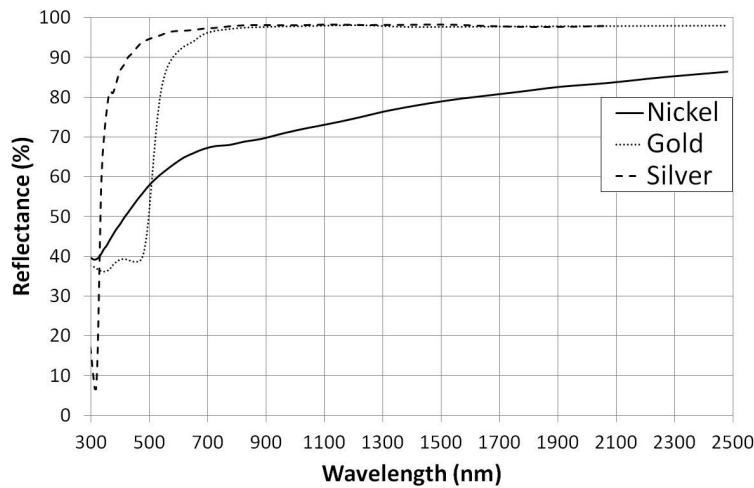

Figure 3: Reflectance at an air-metal interface for nickel, gold and silver

There are a number of approaches to calculating absorption in thin films, the most straightforward way of analysing this situation is using a method known as the Transfer Matrix Method[15]. Here the transmission and reflection coefficients for each interface and the uniform material section are combined to give an overall transmission and reflection response for the complete layer structure.

There are no assumptions made here about the thickness of the layer with respect to wavelength as is required in some of the analytic approaches that can be used.

Figure 4 shows the Transmittance, *T* and Reflectance, *R* for nickel, gold and silver for 3 different films thicknesses, 10, 25 and 50nm. These have been checked against online available thin film calculator at for example www.luxpop.com

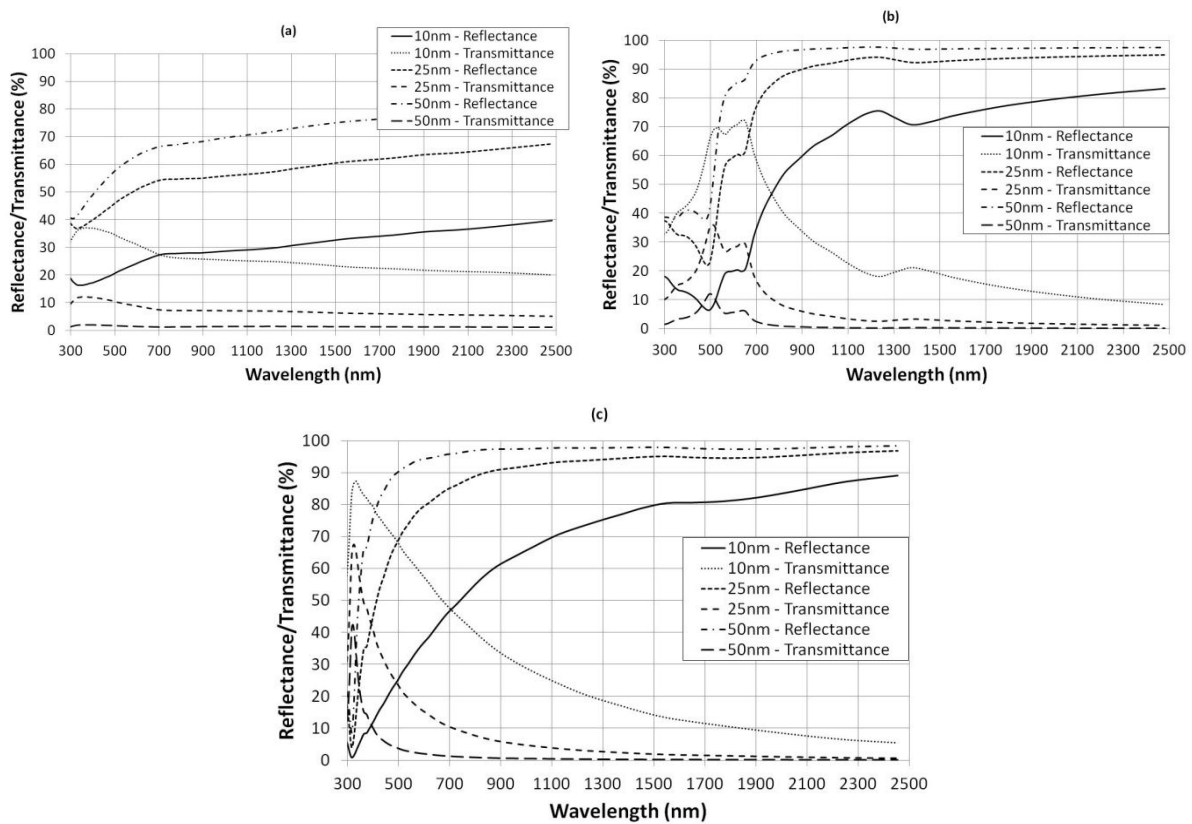

Figure 4: TMM models of *R* and *T* for 10nm, 25nm and 50nm (a) nickel, (b) gold and (c) silver films in air

In the case of silver we can compare our results to [11], which shows a transmittance and reflectance of a 9nm thick film at a wavelength of 1000nm to be ~30% and 55% respectively and this is very close to our modelled values. In [11] the layer is on glass rather than in air as in our case, this will only make a small difference to these results. For nickel, reference [12] shows a transmittance of ~40% at 400nm which is again close to our modelled value. These results show that our data for real and imaginary dielectric constant result in realistic values for transmittance and reflectance for thin films.

For solar thermal applications we are interested in Absorptance, A, which in this simple one dimensional model with no scattering can be calculated directly from Transmittance and Reflectance using: A=1-T-R. The calculated Absorptance is shown in Figure 5 for different metal thickness:

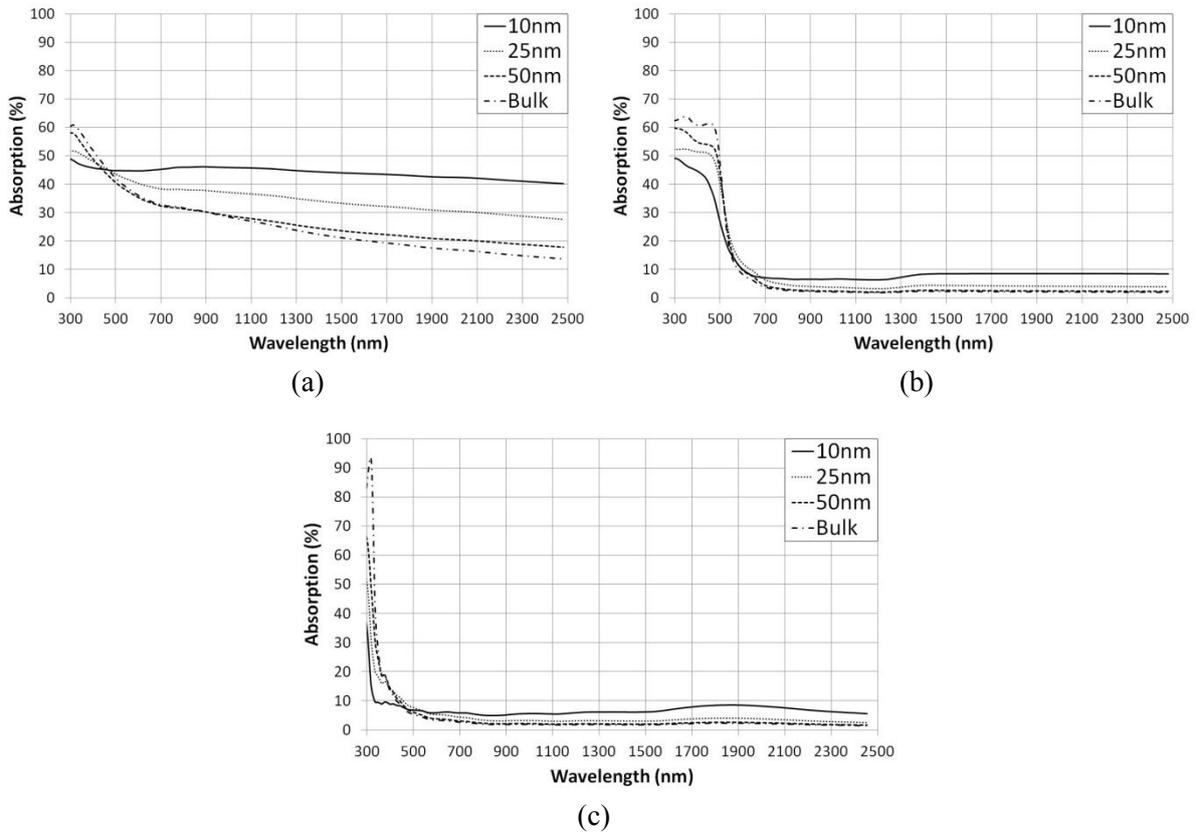

Figure 5: TMM models of absorption of bulk, 50nm, 25nm and 10nm of (a) nickel, (b) gold and (c) silver films

These results show the absorption can strongly depend on the thickness and wavelength. In each case it can be seen that there is a cross over point for the 10nm thick case where its absorption is greater than the 25nm case. In the case of nickel with its very strong absorption across the solar spectrum this could give rise to a significant increase in overall absorption.

In order to quantify the total absorption we have integrated the absorption across this spectral range and normalised to 100% absorption at all wavelengths, the data is shown in Figure 6. The figure shows that for silver there is a gradual increase with increasing thickness as one would expect. For gold there is a very slight local maximum around 15nm with a bulk value of ~31%. In the case of nickel we can observe a pronounced peak at 12.5nm reaching a maximum value of 46%, compared to a bulk value of 40%. These results are in line with the general theory presented in [16] and we believe this is one of the first examples of a detailed study of these effects in nickel films.

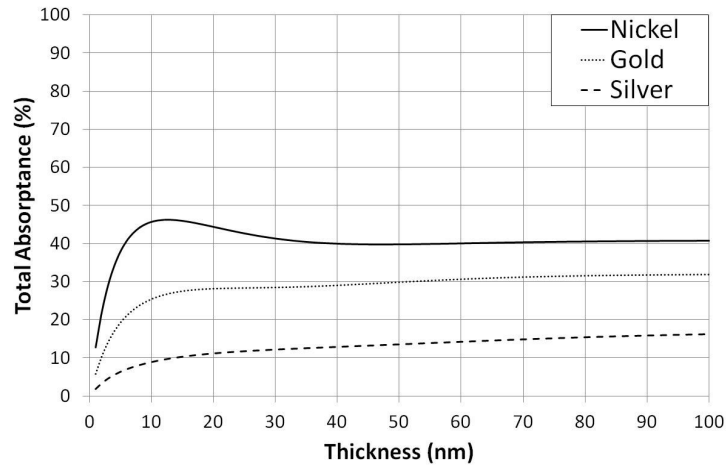

Figure 6: Total integrated solar absorption for nickel, gold and silver films of different thicknesses

**Conclusion**

This paper has studied absorption in ultra-thin layers of nickel, gold and silver. It is shown that nickel possesses optical properties that make it ideal for use in solar thermal and solar thermionic applications and that there is an optimum thickness for maximising absorption across the solar spectrum of ~10-13nm. Experimental studies are now under way to confirm these modelled results. Some points to note are (i) we have assumed perfectly flat films, which will not be the case in reality, (ii) we have also not looked at the angular dependence of absorption which is very important for solar thermal applications. These aspects will be addressed in future publications. This initial modelling study is a precursor to a full 3D Finite Difference Time Domain study of micro and nanostructured interfaces where the enhancement shown here could be dramatically increased.

**Acknowledgements**
The Authors would like to acknowledge useful discussions with Dr Jay Sarma

**References**
[1] C. G. Granqvist, "Spectrally Selective Coatings for Energy Efficiency and Solar Applications", Physica Scripta. Vol. 32,401-407, 1985.
[2] H. Naito *et al*, "Development of a solar receiver for a high-efficiency thermionic/thermoelectric conversion system", Solar Energy Vol. 58, No. 4-6, pp. 191-195, 1996.
[3] T.L. Martin, K.M. O'Donnell, H.Shiozawa, C. E. Giusca, N.A. Fox, S. R. P. Silva and D. Cherns., "Lithium monolayers on single crystal C(100) oxygen-terminated diamond", 2010 MRS Fall Meeting. MRS Proceedings (2011), 1282 : doi: 10.1557/opl.2011.449
[4] Q-C. Zhang and D.R.Mills, 'New cermet film structures with much improved selectivity for solar thermal applications', Appl. Phys. Lett, 60(5), 545-547,(1992).
[5] S. Y. Lin, J. Moreno, and J. G. Fleming, "Three-dimensional photonic-crystal emitter for thermal photovoltaic power generation" Applied Physics Letters Vol. 83, No. 2 (2003)
[6] Y. Cui *et al*. "A thin film broadband absorber based on multi-sized nanoantennas", Appl. Phys. Lett. 99, 253101 (2011)
[7] *Optical Metamaterials: Fundamentals and Applications*,Wenshan Cai, Vladimir Shalaev,, Springer 2010
[8] J Bhattacharya1, N. Chakravarty "A photonic-plasmonic structure for enhancing light absorption in thin film solar cells", Appl. Phys. Lett. 99, 131114 (2011)
[9] M. Thiel, J. Fischer, G. von Freymann, and M. Wegener, "Direct laser writing of three-dimensional nanostructures using a continuous-wave laser at 532nm", Applied Physics Letters 97, 221102 (2010)
[10] H.A.Atwater, "The Promise of Plasmonics" Scientific American, April 2007

[11] C. G. Granqvist, "Transparent conductors as solar energy materials: A panoramic review", Solar Energy Materials & Solar Cells 91 (2007) 1529–1598

[12] D. S. Ghosh, L. Martinez, S. Giurgola, P. Vergani, and V. Pruneri, "Widely transparent electrodes based on ultrathin metals", Optics Letters, Vol. 34, No. 3, February 2009

[13] *Plasmonics: Fundamentals and Applications,* S.Maier, Springer, 2007

[14] www.filmetrics.com/refractive-index-database

[15] *Principles of optics: electromagnetic theory of propagation, interference and diffraction of light* Born, M.; Wolf, E., Oxford, Pergamon Press, 1964.

[16] E.F.C. Driessen, F.R. Braakman, E.M. Reiger, S.N. Dorenbos, V. Zwiller, and M.J.A. de Dood, "Impedance model for the polarization-dependent optical Absorption of superconducting single-photon detectors", Eur. Phys. J. Appl. Phys. 47, 10701 (2009)

[17] C. Hagglund, S. P. Apell, and B. Kasemo, "Maximized Optical Absorption in Ultrathin Films and Its Application to Plasmon-Based Two-Dimensional Photovoltaics", Nano Lett. 2010, 10, 3135–3141

[18] *Optical Properties of Solids,* p. 49. Wooten, Frederick, Academic Press (1972).